\documentstyle[preprint,pra,aps,eqsecnum,amsfonts]{revtex}

\tightenlines

\begin{document}

\draft
\title{Lattice representations of Penrose tilings of the plane}
\author{Matthias W. Reinsch}
\address{Department of Physics, University of California, Berkeley, California
94720, {\tt reinsch@uclink4.berkeley.edu}}
\date{\today}
\maketitle

\begin{abstract}
Two-, three- and four-dimensional representations of Penrose
tilings of the plane are described.  The vertices that occur in these
representations lie on lattices.  Symmetries and methods of visualizing these
representations are discussed.  The question of efficiently storing
the information necessary to reconstruct a tiling is addressed.

\end{abstract}

\section{introduction}

The subject of Penrose tiles has been studied extensively, and the
concept of quasiperiodicity has found applications in
physics\cite{RMP}.  The definition of the Penrose tiles discussed in
this paper is the same as that of Refs.~\onlinecite{RMP} and
\onlinecite{R2}, but the methods presented here can be applied to
other types of tiles as well.  We consider a tiling of the plane by a
rhombus containing an angle of $36^\circ$ and a rhombus containing an
angle of $72^\circ$; the edges of these two tiles have unit length.
Some authors work with additional matching rules, according to which
certain arrangements of tiles are forbidden.  The methods of this paper
can be applied when matching rules are present as well as when they
are absent.  An example of a tiling
is shown in Fig.~1.  Although there are only finitely many
directions possible for the edges in such a tiling, the vertices do
not lie on any lattice (a lattice is a set of points obtained by
taking all integer linear combinations of a set of linearly
independent vectors).  The subject of this paper is two-, three- and
four-dimensional
representations of Penrose tilings of the plane.  Given a tiling of
the plane, a lattice representation is defined by placing
tiles with four vertices on a lattice, which has a dimensionality
of two, three or four.  The tiles share edges with neighboring tiles in the
same way as in the original tiling of the plane.
Figure~2 shows an example of a three-dimensional
representation of the tiling in Fig.~1.

\section{four-dimensional representation}

We begin by defining a four-dimensional representation of a Penrose 
tiling of the plane.
Such representations are well-known\cite{A1,A2,A3,A4}, but we present a
discussion here because the later sections of this paper make use of this material.
We define $x$ and $y$ axes in the plane in such a 
way that one of the vertices of the tiling is located at the origin, and 
that the angles between edges of tiles and the $x$ axis are multiples 
of $36^\circ$.  Because of the relationships
\begin{eqnarray}
\cos 36^\circ &=& \frac{1+\sqrt{5}}{4} \, , \nonumber \\
\cos 72^\circ &=& \frac{-1+\sqrt{5}}{4} \, ,
\end{eqnarray}
and $\cos 0 = 1$,
the $x$ coordinates of all of the vertices in the tiling can be 
expressed as integer linear combinations of $1/4$ and $\sqrt{5}/4$.  
The $y$ coordinates of all of the vertices in the tiling can be 
expressed as integer linear combinations of $\sin 36^\circ$ and $\sin 
72^\circ$.  Thus the $x$ and $y$ coordinates of any vertex in the 
tiling can be described by four integers $x_1$, $x_2$, $x_3$ and 
$x_4$:
\begin{eqnarray}
x &=& \frac{x_1 + x_2 \, \sqrt{5}}{4} \, , \nonumber \\
y &=& x_3 \, \sin 36^\circ + x_4 \sin 72^\circ \, . \label{eq2}
\end{eqnarray}
The integers $x_1$, $x_2$, $x_3$ and $x_4$ are the coordinates of 
the given vertex in a four-dimensional space.  The set of points in 
I~\hspace{-.07in}R$^4$ with integer coordinates is a lattice.  Not all 
of the points of this lattice are used in our description of a Penrose 
tiling.  Figure~3 shows the coordinates in I~\hspace{-.07in}R$^4$ of 
ten points in the plane.  These are the ten possible displacement 
vectors for a step, as one traces a path along the edges in a tiling of 
the plane.  The four integers corresponding to such a vector give the 
changes in the coordinates in I~\hspace{-.07in}R$^4$ as one moves 
along the edge of a tile.
In the four-dimensional representation, a tile is specified by four 
points in I~\hspace{-.07in}R$^4$; going around the tile along the 
edges involves taking steps in I~\hspace{-.07in}R$^4$ given by the 
numbers in Fig.~3.  Thus we see that a tile in I~\hspace{-
.07in}R$^4$ has displacement vectors along its edges that are 
independent of the location of the original tile in the plane.  What 
does matter is the orientation of the original tile in the plane.  
Different orientations of the same tile in the plane correspond to tiles 
having different shapes in I~\hspace{-.07in}R$^4$.  In this sense, 
the number of types of tiles that occur in the four-dimensional 
representation is greater than the original value of two.

Given the four-dimensional representation of a tiling of the plane, 
the original tiling can be recovered using Eq.~(\ref{eq2}).  This 
corresponds to projecting the four-dimensional tiling onto a
two-dimensional subspace, with suitably defined coordinates in the 
subspace.  Since objects in four dimensions are difficult to visualize, 
it is also of interest to consider projections onto three-dimensional 
subspaces.  This can be done without creating self-intersections of 
the higher-dimensional tiling.  Then a further projection results in 
the original tiling of the plane.  These matters are discussed in the next section.

Rotating a Penrose tiling of the plane by $36^\circ$ results in a tiling 
that has the same set of allowed directions for the edges, so the 
vertices can be described by four integers using the same procedure 
as for the original tiling.  Using Eq.~(\ref{eq2}), one finds that the 
effect of a counterclockwise rotation by $36^\circ$ on the four 
integers identifying a vertex can be described by multiplication by 
the following matrix:
\begin{equation}
T = \frac{1}{4} \left(\matrix{
1 & 5 & -10 & 0 \cr
1 & 1 & 2 & -4 \cr
1 & -1 & 0 & 2 \cr
0 & 2 & 2 & 2 \cr
} \right) \, ,
\label{R36}
\end{equation}
Because of the definition of $T$ as a rotation
in the plane by an angle of 
$36^\circ$
(one tenth of a revolution), we have the following identities for higher powers of $T$:
\begin{eqnarray}
T^5 &=& -I \, ,\nonumber \\
T^{10} &=& I \, . \\
\end{eqnarray}
Here $I$ denotes the $4\times4$ identity matrix.
The matrix $T$ is not orthogonal (that is, $T \, T^t$ is not the identity 
matrix), but
it does possess the property
\begin{equation}
T \, M \, T \, M = I \, ,
\label{TM}
\end{equation}
where the matrix $M$ is defined to be
\begin{equation}
M = \left(\matrix{
-1 & 0 & 0 & 0 \cr
0 & -1 & 0 & 0 \cr
0 & 0 & 1 & 0 \cr
0 & 0 & 0 & 1 \cr
} \right) \, .
\label{M}
\end{equation}
The reason Eq.~(\ref{TM}) is true is that the action of $M$ is the 
same as a reflection about the $y$ axis in the original
two-dimensional space, and the conjugation of a rotation by a reflection in 
I~\hspace{-.07in}R$^2$ is the inverse of the rotation.

The result of multiplying a column vector of four integers that 
represent a
vertex in a Penrose tiling
by the matrix $T$ must again be a column vector of four integers, for 
reasons explained above.  This statement yields some constraints on 
the possible coordinates in I~\hspace{-.09in}R$^4$ for vertices in a 
Penrose tiling of the plane.  For each row in the matrix $T$ we have a 
constraint.  For example, from the first row, we see that $x_1 + 5 x_2 
-10 x_3$
must be a multiple of four.  Because the $x_i$ are integers, this 
statement is equivalent to the statement that $x_1 + x_2 + 2 x_3$ 
must be a multiple of four.
Another way to prove this statement is to look at the coordinates of 
the fundamental displacement vectors shown in Fig.~3.  For each of 
these steps, $x_1 + x_2 + 2 x_3$ is a multiple of four, so $x_1 + x_2 + 
2 x_3$ must be a multiple of four for arbitrary combinations of these 
steps.  Looking at the other rows in the matrix $T$ we obtain further 
statements about sets of four integers that represent a
vertex in a Penrose tiling.  Some of this information is redundant.  
The information may be summarized as
\begin{eqnarray}
x_1 + x_2 + 2 x_3 &=& 0 \,\, mod \,\, 4 \, , \nonumber \\
x_2 + x_3 + x_4 &=& 0 \,\, mod \,\, 2 \, .
\end{eqnarray}
If we define new coordinates according to
\begin{eqnarray}
x_1^\prime &=& \frac{x_1 + x_2 + 2 x_3}{4} \, , \nonumber \\
x_2^\prime &=& \frac{x_2 + x_3 + x_4}{2} \, , \nonumber \\
x_3^\prime &=& x_3 \, , \nonumber \\
x_4^\prime &=& x_4 \, ,
\end{eqnarray}
then we may use $x_1^\prime$, $x_2^\prime$, $x_3^\prime$ and 
$x_4^\prime$ as coordinates on I~\hspace{-.1in}R$^4$.  These coordinates have the property that all possible integer values 
are taken on when representing arbitrary finite sums of the 
fundamental displacement vectors.  If complex numbers are used to 
describe points in the plane, these sums become sums of unimodular 
complex numbers of the form $\exp(i \pi n_j / 5)$, where the $n_j$ 
are integers.  The points obtained in this way represent vertices that 
would occur in tilings of the plane, if overlapping tiles were allowed.
The fact that all possible integer values are taken on by the primed 
coordinates follows from the observation that (in the complex 
notation for points in the plane) $\exp(0)$ is represented by 
$(1,0,0,0)$, $\exp(2\pi i/5)+\exp(-2\pi i/5)$ is represented by 
$(0,1,0,0)$, $\exp(4\pi i/5)$ is represented by $(0,0,1,0)$, and 
$\exp(3\pi i/5)$ is represented by $(0,0,0,1)$.

With respect to the primed coordinates, the matrix $T$ becomes
\begin{equation}
T^\prime = \left(\matrix{
1 & 0 & -1 & 0 \cr
1 & 0 & 0 & 0 \cr
1 & -1 & 0 & 1 \cr
0 & 1 & 0 & 0 \cr
} \right) \, .
\end{equation}
The fact that only the numbers 1, -1 and 0 are present in this matrix indicates that an argument 
such as the one presented above for the matrix $T$ will not give any 
constraints on the values of the
new coordinates.

The $x$ and $y$ coordinates of any vertex in the tiling can be 
described by the four integers $x_1^\prime$, $x_2^\prime$, 
$x_3^\prime$ and $x_4^\prime$ as
\begin{eqnarray}
x &=& \frac{4 x_1^\prime - 2 x_2^\prime - x_3^\prime + x_4^\prime
+ (2 x_2^\prime - x_3^\prime - x_4^\prime) \, \sqrt{5}}{4} \, , 
\nonumber \\
y &=& x_3^\prime \, \sin 36^\circ + x_4^\prime \sin 72^\circ \, . 
\label{eq2prime}
\end{eqnarray}
Because of the complexity of this transformation, we prefer to use
the unprimed coordinates, shown in Eq.~(\ref{eq2}).

\section{three-dimensional representations}

In this section we describe three-dimensional representations of
Penrose tilings of the plane.  These have the advantage that they are
easier to visualize than the four-dimensional representation 
described
in the previous section.  Two projections are described.  We call
these the ``$\mu$-projection'' and the
``(1,2)-projection.'' The $\mu$-projection has the advantage
that the original tiling of the plane can be obtained by a simple
projection onto a certain two-dimensional subspace of I~\hspace{-
.07in}R$^3$.  It has
the disadvantage that only the $x$ and $y$ coordinates of the 
vertices
are integers.  The set of $z$ values that occur are not integer
multiples of a basic step.  The (1,2)-projection has the advantage
that the $x$, $y$ and $z$ coordinates of the vertices are all
integers.  The disadvantage is that the original tiling cannot be
recovered by a further simple projection, although it can be 
reconstructed
by a different means, as explained below.

\subsection{The $\mu$-projection}

We define a mapping from I~\hspace{-.07in}R$^4$ to I~\hspace{-
.07in}R$^3$ using the matrix
\begin{equation}
P^{(\mu)} = \left(\matrix{
1 & 0 & 0 & 0 \cr
0 & 1 & 0 & 0 \cr
0 & 0 & \mu \sin\frac{\pi}{5} & \mu \sin \frac{2\pi}{5} \cr
} \right) \, ,
\end{equation}
where $\mu$ is a real number specified below.  (The $\mu$ on the
left-hand side of the equation is not an index.)  Under this mapping, the vertices in
I~\hspace{-.07in}R$^4$ map to points in I~\hspace{-.07in}R$^3$.  
The resulting arrangement of tiles in
I~\hspace{-.07in}R$^3$ can be projected onto the plane 
perpendicular to $(\sqrt{5}, -1, 0)$ to recover the original tiling.  The 
proper choice for the
value of $\mu$ results in a two-dimensional tiling in which the
angles are the same as in the original tiling.  We will use
$\frac{1}{\sqrt{6}} (1, \sqrt{5}, 0)$ and $(0,0,1)$ as orthonormal
basis vectors for the two-dimensional subspace perpendicular to
$(\sqrt{5}, -1, 0)$, and we let $x'$ and $y'$ denote the
coordinates of projected points with respect to this basis.  The reason 
for this
choice of subspace will become clear in the following calculation.  The
values of $x'$ and $y'$ for a point $(x_1, x_2, x_3, x_4)$ in
I~\hspace{-.07in}R$^4$ can be obtained from
\begin{eqnarray}
\left(\matrix{
x' \cr
y' \cr
} \right) &=&
\left(\matrix{
\frac{1}{\sqrt{6}} & \frac{5}{\sqrt{6}} & 0 \cr \noalign{\vskip 5pt}
0 & 0 & 1 \cr } \right) \, 
\left(\matrix{
1 & 0 & 0 & 0 \cr
0 & 1 & 0 & 0 \cr
0 & 0 & \mu \sin\frac{\pi}{5} & \mu \sin \frac{2\pi}{5} \cr
} \right) \, \left(\matrix{
x_1 \cr
x_2\cr
x_3 \cr
x_4 \cr
} \right)
\nonumber \\ 
&=& 
\left(\matrix{
\frac{1}{\sqrt{6}} & \frac{5}{\sqrt{6}} & 0 & 0 \cr \noalign{\vskip 5pt}
0 & 0 & \mu \sin\frac{\pi}{5} & \mu \sin \frac{2\pi}{5} \cr
} \right) \, \left(\matrix{
x_1 \cr
x_2\cr
x_3 \cr
x_4 \cr
} \right)
\nonumber \\ 
&=& 
\left(\matrix{
2 \sqrt{\frac{2}{3}} \, x \cr \noalign{\vskip 5pt}
\mu \, y \cr
} \right) \, ,
\end{eqnarray}
where we have used Eq.~(\ref{eq2}).  Thus, we must take
\begin{equation}
\mu = 2 \sqrt{\frac{2}{3}} \, .
\end{equation}
For arrangements of a finite
number of Penrose tiles, the process of projecting onto the plane
perpendicular to $(\sqrt{5}, -1, 0)$ can be implemented by viewing 
the
three-dimensional representation from a viewpoint in the $(\sqrt{5},
-1, 0)$ direction.  The image obtained in the limit as the distance
from the tiles to the observer becomes infinite (together with
suitable magnification of the image) is the projection onto the plane
perpendicular to $(\sqrt{5}, -1, 0)$.  Figure~5 shows the
three-dimensional representation corresponding to the tiles in the
plane shown in Fig.~4, using the $\mu$-projection.  Figure~6 contains
a cut-out model of the object shown in Fig.~5.  The lower portion is a
frame to support the model.  Its location in the fully assembled 
model
corresponds roughly to the right half of the encompassing box in
Fig.~5.

\subsection{The (1,2)-projection}

The (1,2)-projection is defined by the matrix
\begin{equation}
P^{(1,2)} = \left(\matrix{
1 & 0 & 0 & 0 \cr
0 & 1 & 0 & 0 \cr
0 & 0 & 1 & 2 \cr
} \right) \, ,
\end{equation}
The numbers in the lowest row of this
matrix provide the motivation for the name ``(1,2)-
projection.'' The numerical values in this matrix are close to
those in $P^{(\mu)}$, so the resulting projected tilings are similar
in appearance.  However, since the coordinates $x_3$ and $x_4$ 
occur
in the combination $x_3 + 2 x_4$, it is not possible to obtain $y = x_3 
\, \sin 36^\circ + x_4 \sin 72^\circ$ by simple algebraic operations.  
In spite of this, it is
possible to recover the original tiling from the (1,2)-projection
because the tiles in the projection indicate the types of tiles
(narrow or wide) and their connections (shared edges) in the original
tiling.  Therefore, the information that is necessary to assemble the 
original tiling
is contained in the (1,2)-projection.
As mentioned above, this three-dimensional representation has
the property that the $x$, $y$ and $z$ coordinates of the vertices are
integers.  Figure~2 shows the (1,2)-projection corresponding to Fig.~1.

\section{two-dimensional representations}

It is also possible to create a two-dimensional lattice representation 
of a Penrose tiling of the 
plane.  One way to do this is to
map the vertices in I~\hspace{-.07in}R$^4$ to 
I~\hspace{-.07in}R$^2$ using the matrix
\begin{equation}
\left(\matrix{
\frac{1}{4} & \frac{1}{2} & 0 & 0 \cr \noalign{\vskip 5pt}
0 & 0 & \frac{1}{2} & 1 \cr
} \right) \, ,
\label{35}
\end{equation}
The resulting tiling of the plane (see Fig.~7)
resembles the original tiling, but
the tiles are slightly different.
Also, the set of directions of edges is no longer
invariant under a rotation through an angle of $36^\circ$.
There are three types of narrow
tiles and three types of wide tiles
(modulo reflections about the coordinate axes), corresponding to different
orientations of the tiles in the original tiling.  As above, the types
of tiles (narrow or wide) and their connections to each other (shared
edges) can be identified, so the original tiling can be reconstructed,
but this process is not a simple projection operation.  Although some
of the vertices move quite a bit in this process, nearby vertices are
moved by a similar amount, and no self-intersections of the pattern 
are
created.  This can be seen by applying the transformation to the
points in Fig.~3.  These displacement vectors occur along the edges of
the tiles.  The point (1,0) [which is represented by (4,0,0,0) in the
four-dimensional representation] gets mapped to (1,0); the point 
$(\cos 36^\circ,
\sin 36^\circ)$ [which is represented by (1,1,1,0) in the
four-dimensional
representation] gets mapped to $(\frac{3}{4}, \frac{1}{2})$, etc.
Thus, the shapes of the tiles are not changed drastically, and the
identities and arrangement of the original tiles can be read off of
the transformed tiling.  This is as in the case of the
(1,2)-projection above.  Another way to say this is that if a vertex
is selected in Fig.~7, its original coordinates can be found by
choosing an arbitrary path back to the origin along edges in the
diagram.  The steps in this path identify unit vectors which must be summed
to obtain the coordinates of the vertex in Fig.~1.

The description of a tiling can be further simplified by removing the
edges of the tiles from the list of information that is recorded about
the tiling.  It is sufficient to record only the locations of the vertices
of the tiles on the two-dimensional lattice, as shown in Fig.~8.  Only
certain types of displacement vectors can occur as edges.  These were
mentioned in the preceding paragraph.  Thus, given the information in
Fig.~8, the edges can be reconstructed.

The description of a tiling using only a record of the locations of
the vertices in the two-dimensional lattice representation, with no
further information about the vertices (such as which vertices belong
to the same tile), is a description using a two-dimensional array of
bits (binary digits).  At each lattice point, a tiling vertex can either
be present or absent.  Not all arrays of bits represent tilings, however.
Some displacements between vertices are forbidden.  It must be possible
to reconstruct a tiling from the bits, with all of the edges belonging
to allowed tiles.

The two-dimensional representation provides a simple description of a
tiling using integers.  One advantage that it has is that it is
``random access.''  It is possible to view the representation of an
arbitrary patch of the tiling, using only information about that patch.
Some other descriptions using integers do not have this property.  For
example, if a history of assembly is recorded (a sequence of tile types,
edge numbers and orientations) then more data must be accessed to view
a patch of the tiling.  The two-dimensional representation is also
useful for certain types of computer tiling programs and tile-overlap
detection.

It should be noted that the two-dimensional lattice representation is
not the same as simply superimposing a grid over the original tiling
and moving the vertices to the nearest grid point.  This would result
in different shapes representing a given orientation of a tile,
depending on how it was positioned relative to the grid.  In the
two-dimensional representation, a given orientation of a tile is always
represented by the same shape, regardless of its position.

The two-dimensional representation also provides an interesting distance
measure to work with when tiling the plane.  This is the Euclidean metric
in the lattice space.  Because the original tiling is not obtained by a
simple projection, this metric is difficult to describe in the original
two-dimensional space.  Figure 9 shows a tiling generated by a simple
algorithm using the two-dimensional representation (basically placing
new tiles as close as possible to the origin).  The subjects of
quasicrystallography and the spiral of Archimedes occur in the
literature\cite{A5}.

\section{conclusion}

The description of Penrose tilings of the plane using
lattices has a variety of applications, 
some of which are discussed in this paper and the references.
As a further application, we describe in the appendix
an algorithm for checking for tile
overlap using calculations with the integers that
occur in the four-dimensional
representation.  In this paper we have shown that lower-dimensional
lattice descriptions of Penrose tilings of the plane are possible,
but the reconstruction process becomes more complicated.
Such representations may be of use
in analyzing tilings, and they provide a simple means of describing
a tiling.

An open question is the large-scale structure and curvature of the
higher-dimensional representations, for tilings of the entire plane.
For example, what would Fig.~2 look like as the number of tiles
becomes very large?  Another question is what the most efficient way to
describe a tiling is.  The most efficient method presented in this paper
is the two-dimensional array of bits.  Looking at Fig.~8, it is clear that
some compression of this data is possible.  Furthermore, if one of the bits
is made unknown, it is still possible to reconstruct the tiling.  Thus,
further improvements in efficiency are possible.

\appendix

\section{Checking for tile overlap using integer math}

The subject of this appendix is an algorithm for checking for tile
overlap using calculations with integers.  Additional information,
such as vertex-vertex contact, is also given.  We will consider here
the case of a narrow tile with its long diagonal at an angle of
$54^\circ$ to the $x$-axis and a wide tile with its long diagonal at
an angle of $36^\circ$ to the $x$-axis.  Other cases can be treated in
a similar manner.  Since the orientations of the tiles are fixed in
this discussion, we need only introduce variables to describe the
relative positioning of the tiles.  We let the four integers $x_1$,
$x_2$, $x_3$ and $x_4$ describe the displacement vector from the
lower-left vertex of the narrow tile to the lower-left vertex of the
wide tile.  It is useful to think of the lower-left vertex of the
narrow tile as being at the origin.  Then the location of the wide
tile is determined by $x_1$, $x_2$, $x_3$ and $x_4$.  As we move the
wide tile around the plane (without rotating it), some of the
positions will have contact between the tiles; others will have no
contact.  The set of positions for the lower-left vertex of the wide
tile for which there is contact is a hexagon.  Opposite sides of the
hexagon are parallel, but the hexagon is not regular.  The second set
of Mathematica instructions below draws a diagram to go along with this
discussion.  For each edge of the hexagon, we define a line by
extending the edge infinitely in both directions.  The side of the
line on which a given test point is located is determined by the sign
of the dot product of a normal vector to the line and the vector
difference between the test point and a point on the line.  We choose
the normal vector to point to the inside of the hexagon.  Because the
normal vector is at $90^\circ$ to the line, $\sin 36^\circ$ appears in
its $x$-component, as may be seen from Eq.~(\ref{eq2}) and the fact
that $\sin 72^\circ = \sin 36^\circ (1+\sqrt{5})/2$.  A simplification
that therefore occurs is that an overall factor of $\sin 36^\circ$ may
be dropped in computing the above-mentioned sign.  The problem reduces
to finding the sign of numbers of the form $a + b \sqrt{5}$, which can
be done using integer math (see {\tt sign[\,]}, below).  A test point
strictly inside the hexagon will have all six signs positive.  If at
least one sign is negative, the point is strictly outside the hexagon.
If exactly one of the signs is zero (and the others positive), the
point is on one of the edges of the hexagon, and it is not one of the
endpoints.  This means we have vertex-edge or partial edge contact
between the tiles.  If two of the signs are zero (and the others
positive), the test point is at one of the vertices of the hexagon.
This means the tiles have vertex-vertex contact.  Two of the vertices
of the hexagon represent perfect edge contact.  These are tested for
at the beginning of the program {\tt check[\,]}.

The program {\tt check[\,]} calculates the six signs and returns a
character string describing the overlap.  The second set of
Mathematica instructions below draws a figure showing how the plane is
divided up.  The narrow tile with its lower-left vertex at the origin
is shown.  The wide tile is located at an arbitrary position.  The
dots summarize the information obtained from {\tt check[\,]} at some
representative points.  These include special points, such as the two
points for which we have perfect edge contact (large filled circles),
points which indicate vertex-vertex contact (medium filled circles),
and vertex-edge or partial edge contact (medium unfilled circles).
Most of the points in the figure are small circles.  The unfilled ones
are points that indicate overlap with nonzero area, and the filled
ones are points that indicate no overlap at all.

\newpage

\begin{verbatim}
(* ========= check for overlapping tiles, using integer math:  ========= *)

(* the sign of a + b Sqrt[5] : *)
sign[a_, b_] := Switch[Sign[a] Sign[b], 1, Sign[a], 0,
     If[a == 0, Sign[b], Sign[a]], -1, Sign[a] Sign[a^2 - 5 b^2]]

(* some parameters: *)
Evaluate[Table[b[i, j], {i, 3},{j, -1, 1, 2}]] =
     {{{-1, -1}, {3, 1}}, {{-4, 0}, {8, 0}}, {{-8, 0}, {4, 4}}}

(* check for overlap: *)
check[x1_, x2_, x3_, x4_] := (
     If[{x1, x2, x3, x4} == {1, 1, 1, 0} || {x1, x2, x3, x4} == {-4,0,0,0},
     Return["perfect edge contact"]];
     p[1] = {2 x3 + x4, x4};
     p[2] = {-x1 + x3 + 3 x4, -x2 + x3 + x4};
     p[3] = {-x1 - 5 x2 - 2 x3 + 4 x4, -x1 - x2 + 2 x3}; signs =
     Sort@Flatten@Table[-j Apply[sign, p[i] - b[i,j]], {i, 3}, {j,-1,1,2}];
     If[signs[[1]] == -1, Return["no contact"]];
     If[signs[[2]] ==  0, Return["vertex-vertex contact"]];
     If[signs[[1]] ==  0, Return["vertex-edge or partial edge contact"]];
     "nonzero-area overlap" )

(* ============== The instructions below draw the figure. ============== *)

f["perfect edge contact"]                 = {Disk, .1}
f["no contact"]                           = {Disk, .025}
f["vertex-vertex contact"]                = {Disk, .05}
f["vertex-edge or partial edge contact"]  = {Circle, .05}
f["nonzero-area overlap"]                 = {Circle, .025}

v1 = {c1, s1} = {Cos[Pi/5], Sin[Pi/5]}
v2 = {c2, s2} = {Cos[2 Pi/5], Sin[2 Pi/5]}

DisksCircles = {};  Do[x = N[(x1 + x2 Sqrt[5])/4]; y = N[x3 s1 + x4 s2];
     If[Abs[x] < 2 && Abs[y] < 2, temp = f[check[x1, x2, x3, x4]];
     AppendTo[DisksCircles, temp[[1]][{x,y}, temp[[2]]]]],
     {x1, -5, 5}, {x2, -3, 3}, {x3, -2, 2}, {x4, -2, 2}]

Show[Graphics[{DisksCircles, Line[{{0, 0}, v1, v1 + v2, v2, {0, 0}}],
     Line[Map[# + {c1 - 1/2, -s2 - s1}&,
     {{0, 0}, {1, 0}, {1, 0} + v2, v2, {0, 0}}]]}],
     AspectRatio -> Automatic, Axes -> True]

\end{verbatim}

\newpage

FIGURE CAPTIONS

\vspace{.25 in}

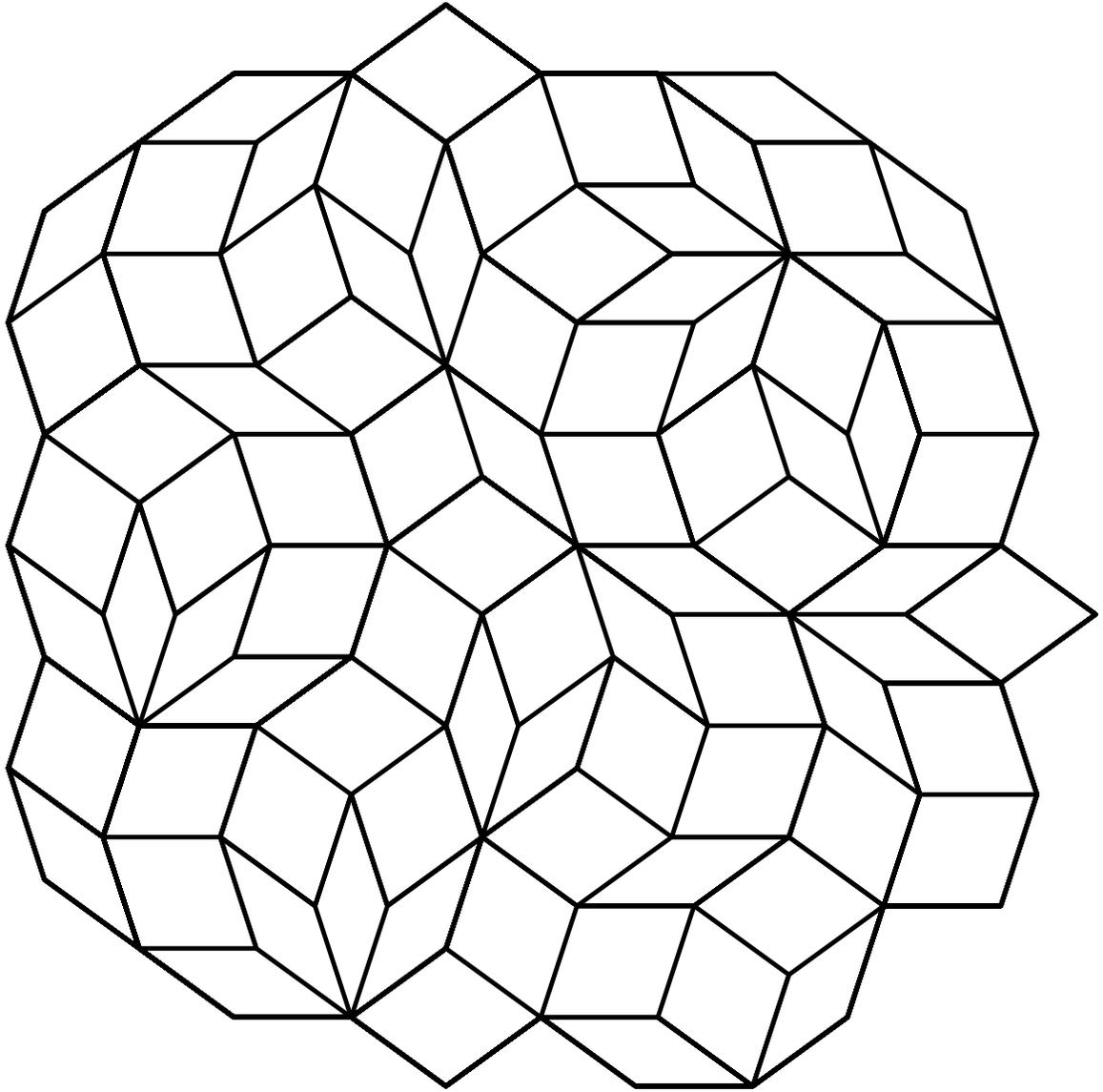
Figure 1:

\vspace{.1 in}

An arrangement of Penrose tiles in the plane.  The vertices of the
tiles do not lie on a lattice.

\vspace{.25 in}

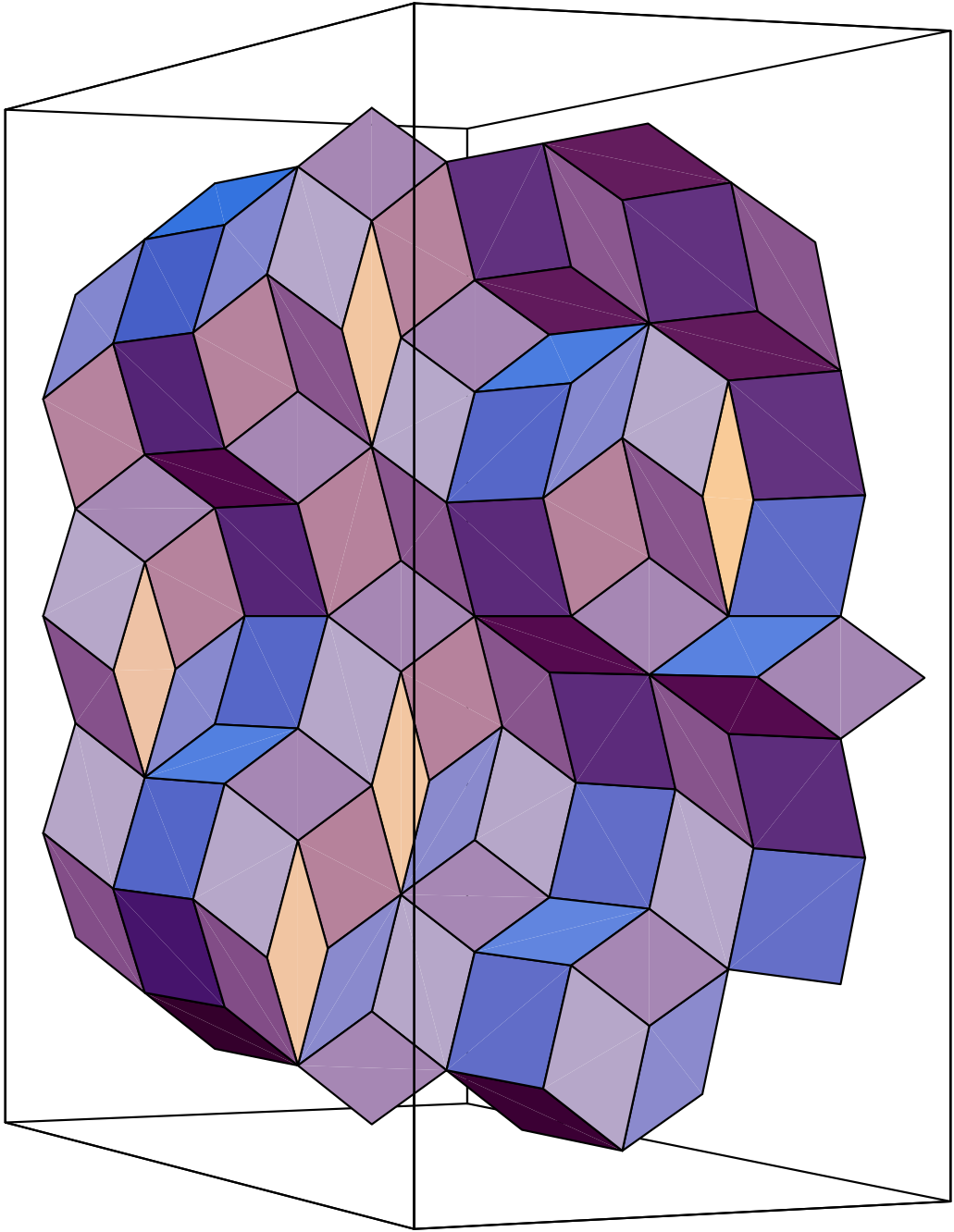
Figure 2:

\vspace{.1 in}

A three-dimensional representation of the tiling shown in Fig.~1.  The $x$, $y$ and $z$ coordinates of the vertices are integers.

\vspace{.25 in}

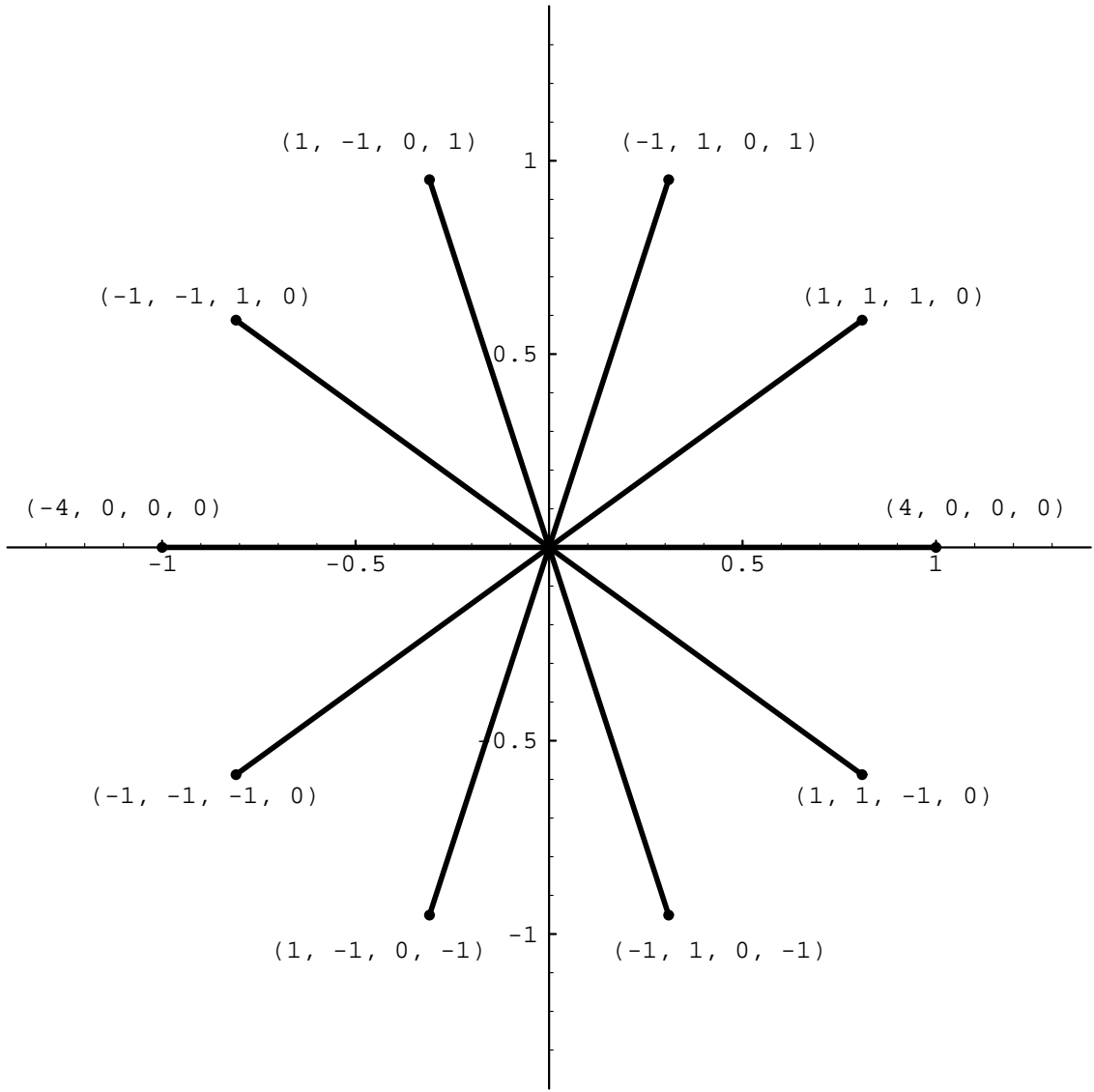
Figure 3:

\vspace{.1 in}

The fundamental displacement vectors in the plane, and their coordinates in I~\hspace{-.07in}R$^4$

\vspace{.25 in}

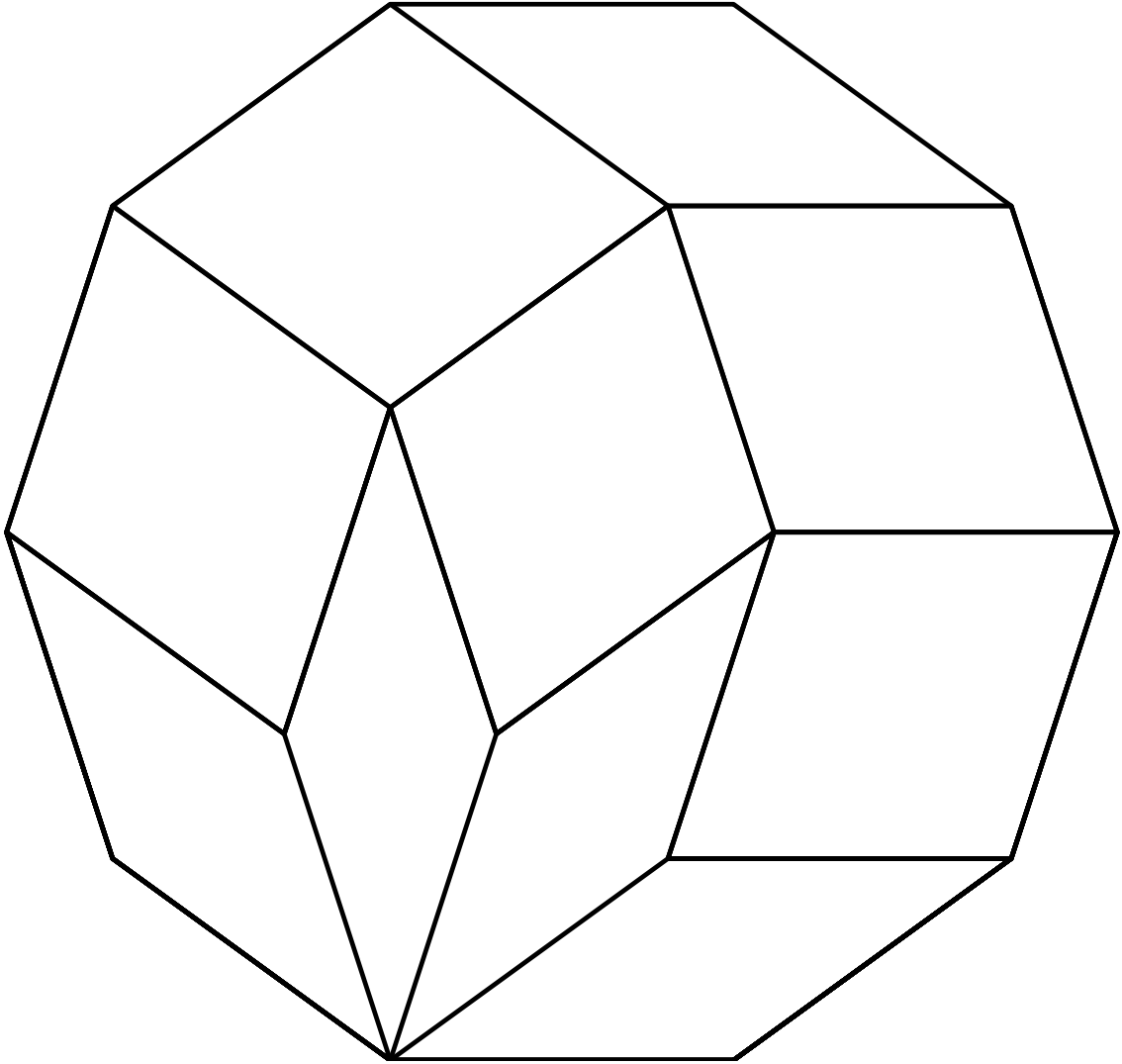
Figure 4:

\vspace{.1 in}

The arrangement of Penrose tiles referred to in Figs.~5 and 6.

\vspace{.25 in}

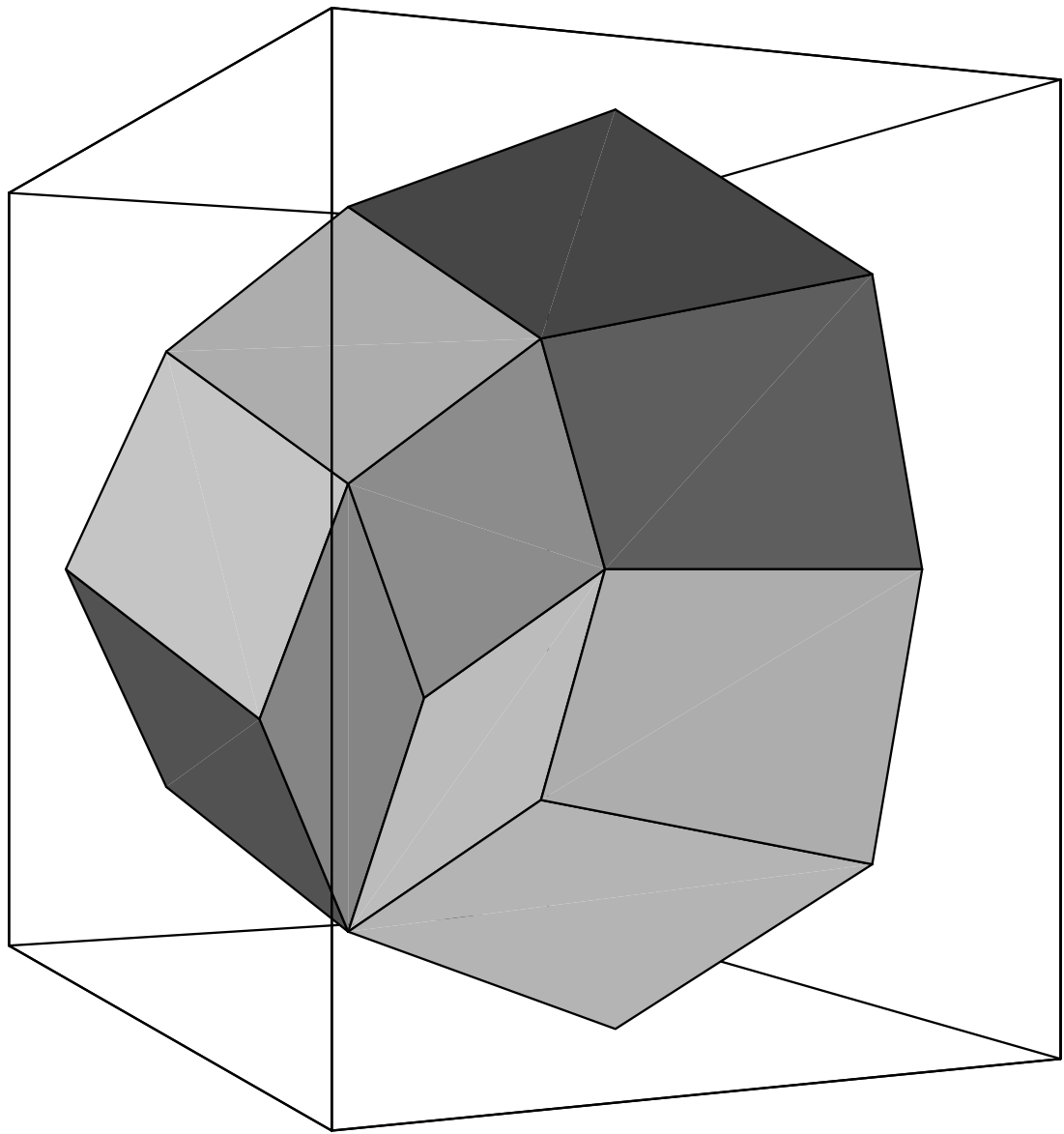
Figure 5:

\vspace{.1 in}

The $\mu$-projection of the tiling shown in Fig.~4.

\vspace{.25 in}

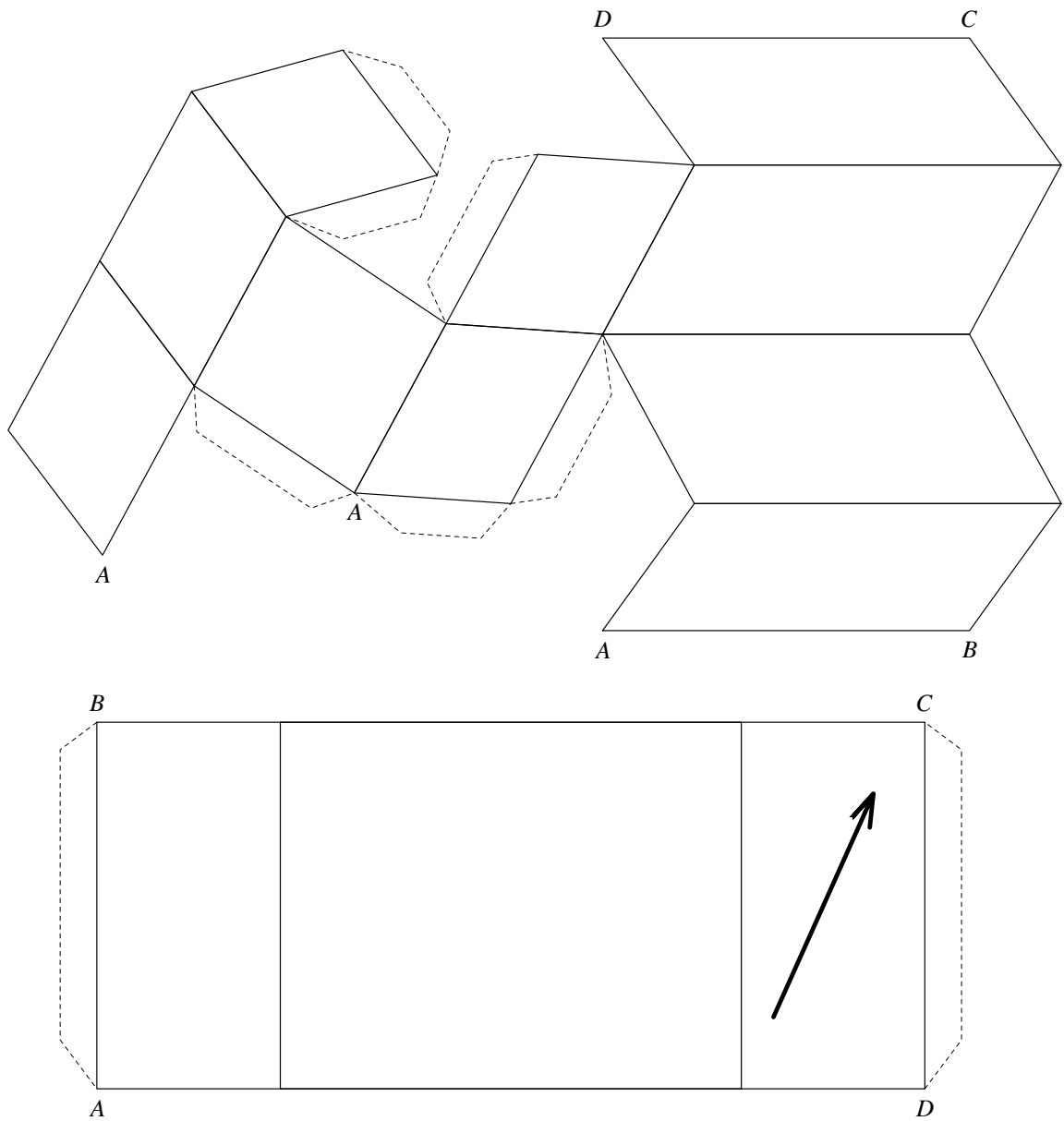
Figure 6:

\vspace{.1 in}

A three-dimensional cut-out model of the representation shown in
Fig.~5.  The letters indicate points that come together to a single
point in the assembled model.  For example, the four points labeled
``A'' come together in the final assembly.  When viewed from a
distance in the direction of the arrow, the model looks like the image
shown in Fig.~4.  Illuminating the model from the left helps to
eliminate distracting shadows.  Further information about the assembly
of the model is given in the text.

\vspace{.25 in}

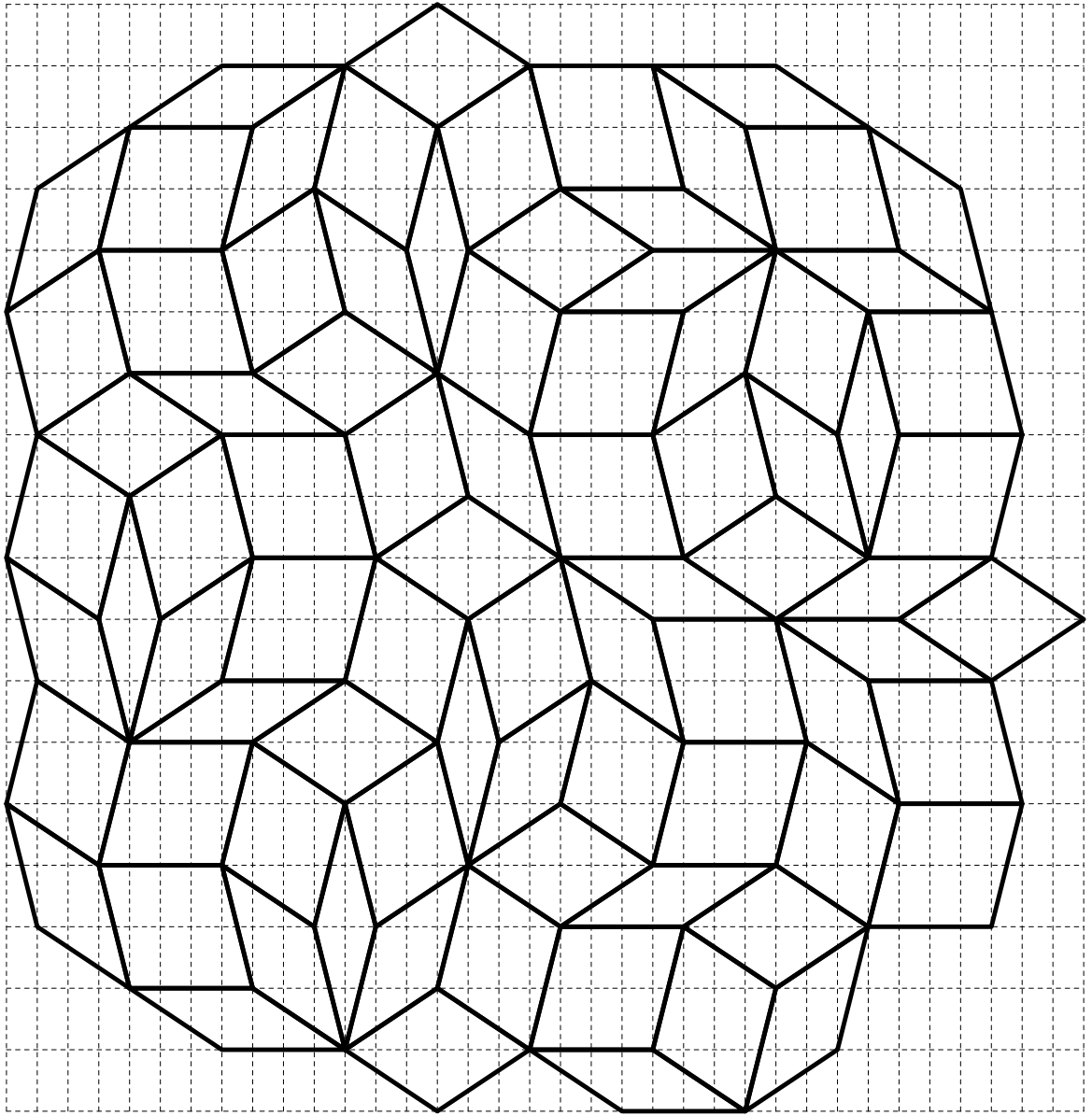
Figure 7:

\vspace{.1 in}

A two-dimensional lattice representation of the tiling shown in Fig.~1.

\vspace{.25 in}

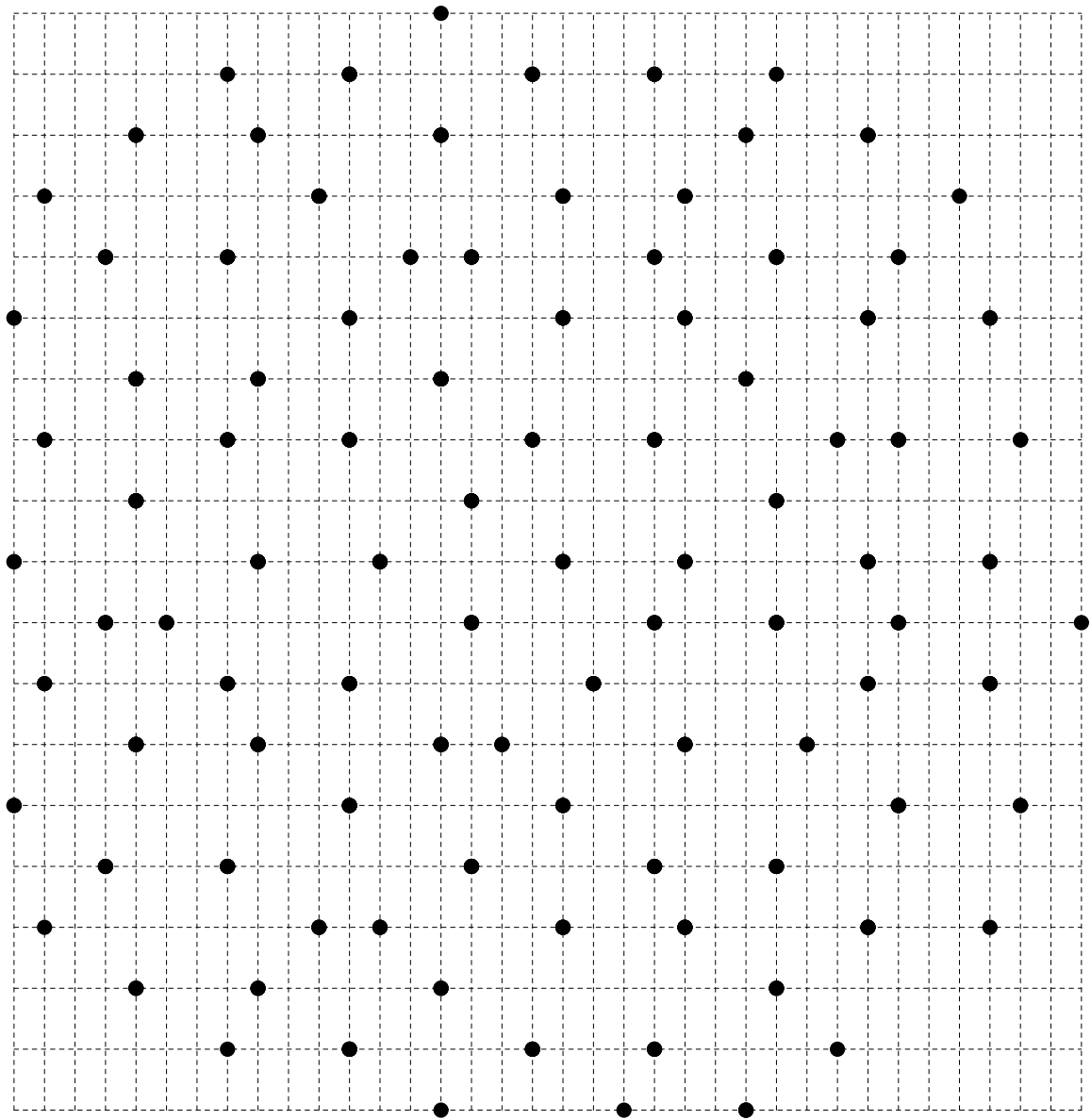
Figure 8:

\vspace{.1 in}

A representation of the tiling shown in Fig.~1 as a two-dimensional array
of bits.  The original tiling can be reconstructed from this information alone.

\vspace{.25 in}

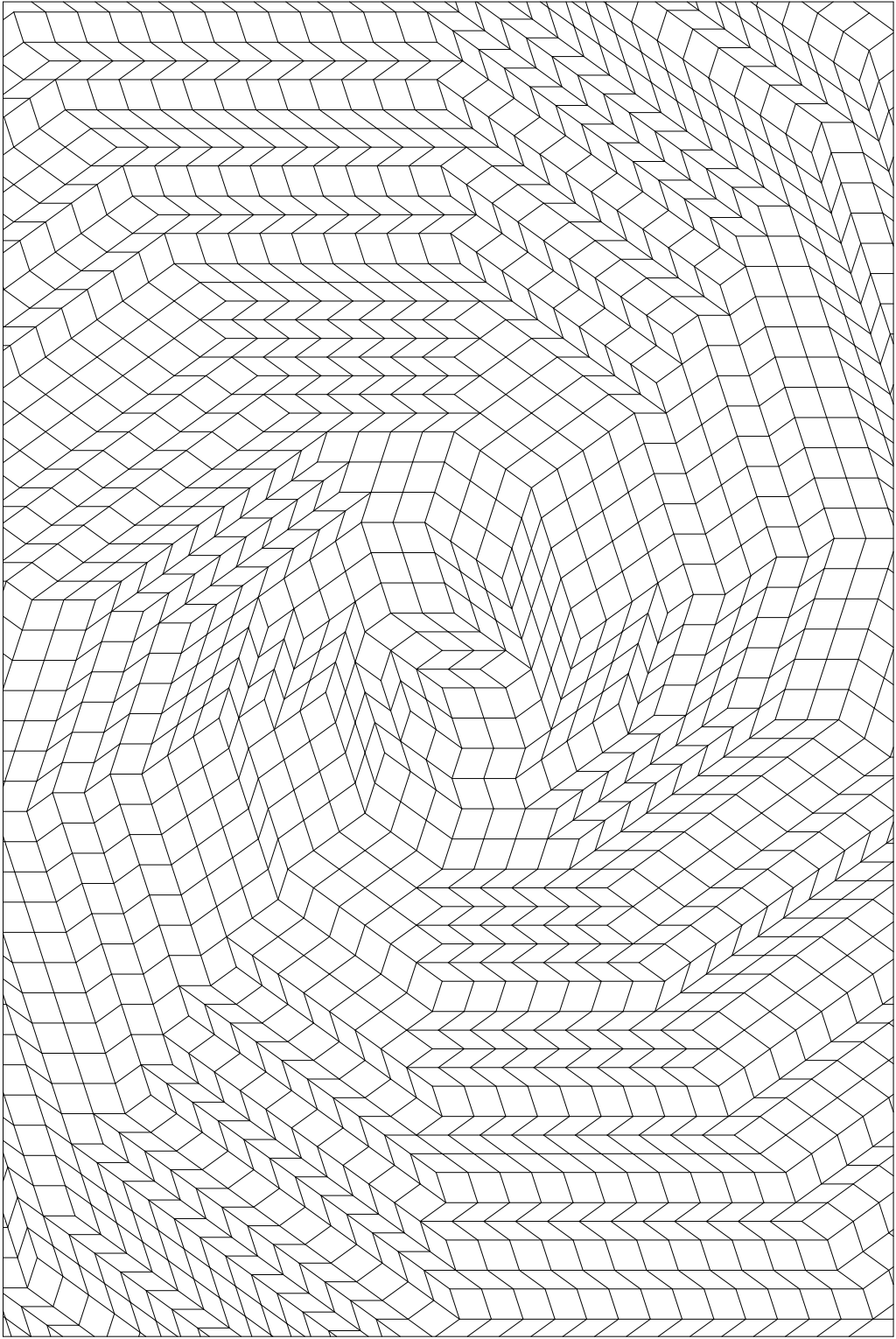
Figure 9:

\vspace{.1 in}

An arrangement of Penrose tiles generated by a simple algorithm using the
two-dimensional lattice representation.

\end{document}